\documentclass[suppldata]{interact}

\usepackage{epstopdf}
\usepackage[caption=false]{subfig}
\usepackage[natbibapa,nodoi]{apacite}
\usepackage{diagbox}
\theoremstyle{plain}

\theoremstyle{definition}

\theoremstyle{remark}

\begin{document}


\title{Mood of India During Covid-19 - An Interactive Web Portal Based on Emotion Analysis of Twitter Data}

\author{
\name{Akhila Sri Manasa Venigalla\textsuperscript{a}, 
Dheeraj Vagavolu\textsuperscript{b} and Sridhar Chimalakonda\textsuperscript{c}}
\affil{Research in Intelligent Software \& Human Analytics (RISHA) Lab\\ Indian Institute of Technology, Tirupati, India\\ \textsuperscript{a}cs19d504@iittp.ac.in
\textsuperscript{b}cs17b028@iittp.ac.in
\textsuperscript{c}ch@iittp.ac.in}
}

\maketitle

\begin{abstract}
The severe outbreak of \textit{Covid-19} pandemic has affected many countries across the world, and disrupted the day to day activities of many people. During such outbreaks, understanding the emotional state of citizens of a country could be of interest to various organizations to carry out tasks and to take necessary measures. Several studies have been performed on data available on various social media platforms and websites to understand the emotions of people against many events, inclusive of \textit{Covid-19}, across the world. Twitter and other social media platforms have been bridging the gap between the citizens and government in various countries and are of more prominence in India. Sentiment Analysis of posts on twitter is observed to accurately reveal the sentiments. Analysing real time posts on twitter in India during \textit{Covid-19}, could help in identifying the mood of the nation. However, most of the existing studies related to \textit{Covid-19}, on twitter and other social media platforms are performed on data posted during a specific interval. We are not aware of any research that identifies emotional state of India on a daily basis. Hence, we present a web portal that aims to display mood of India during \textit{Covid-19}, based on real time twitter data. This portal also enables users to select date range, specific date and state in India to display mood of people belonging to the specified region, on the specified date or during the specified date range. Also, the number of \textit{Covid-19} cases and mood of people at specific cities and states on specific dates is visualized on the country map. As of May 6 2020, the web portal has about 194370 tweets, and each of these tweets are classified into seven categories that include six basic emotions and a neutral category. A list of Trigger Events are also specified, to allow users to view the mood of India on specific events happening in the country during \textit{Covid-19}. 
\end{abstract}

\begin{keywords}
Covid-19; Twitter; Emotion; India
\end{keywords}

\section{Introduction}
\label{intro}
Covid-19 pandemic has been severely affecting many countries across the world. The severity of the pandemic is growing very fast across the world. In India, the number of \textit{Covid-19} cases have greatly increased, from 724 cases to around 29K cases, in a span of one month, from 28 March, 2020 to 28 April, 2020. The total number of deaths reported in the country have increased from 17 on 28 March, 2020 to 934 on 28 April, 2020. There has been a growth rate of more than 100\%, both in view of number of cases and number of deaths in the country, in a span 10 days. Increased number of cases being reported, from around 14K to around 29K and deaths from 480 to 934, indicate the rapid growth of the pandemic in the country (from 18 April, 2020 to 28 April, 2020). It has been observed that the sudden outrages of such pandemics affect public emotion and result in either constructive or destructive behavioural changes [\cite{hou2020assessment}]. Human behavior has been observed to play an important role in either controlling or scaling up spread of a disease, and is more prominent in case of highly contagious pandemics [\cite{siegrist2014role}]. The most common emotions witnessed among the people during several pandemics and crisis situations are fear and anger [\cite{lin2014have}] [\cite{vaughan2009effective}]. Though fear is observed as a basic instinct during the onset of unexpected situations involving danger, researchers also warn that excessive fear results in anxiety disorders, intensifying psychiatric disorders [\cite{garcia2017neurobiology}]. Also, such psychiatric disorders and anxiety might lead to cardiovascular disorders among the vulnerable population [\cite{shin2010neurocircuitry}]. Mental health should hence be considered as an important part of pandemic response [\cite{douglas2009preparing}]. Lack of proper guidance to face the pandemic situation might sometimes result in acute fear, leading to self harm intentions including suicides [\cite{mamun2020first}][\cite{shigemura2020public}]. 

Considering the severity of the pandemic in the context of \textit{Covid-19}, it is extremely important to consider the psychological state and motivate the people accordingly, in controlling the spread of \textit{Covid-19}. Several cases of anxiety, stress and panic have been reported in various countries affected with \textit{Covid-19} such s China, Japan and so on [\cite{li2020chinese}][\cite{shigemura2020public}]. Providing appropriate guidance to the population helps them to respond in a better way when contracted with \textit{Covid-19} [\cite{zhao2020chinese}]. Several organizations across the world such as WHO and CDC, have suggested various measures to be taken up to protect and enhance mental state of vulnerable population during pandemics, including \textit{Covid-19} [\cite{cdccovid}][\cite{Whoburden}]. Thus, analysing public emotions against \textit{Covid-19} helps in understanding the perception and preparedness towards the pandemic among the public [\cite{hou2020assessment}]. It helps in broadcasting effective public health messages that are inline with public emotions, which could help the public in taking appropriate measures against \textit{Covid-19} [\cite{van2020using}]. Several countries have issued strict home isolation and quarantine instructions towards battling \textit{Covid-19}. Governments across the world have taken up health initiatives to reduce the negative psychological impact on the population during the period of home confinement and quarantine [\cite{garriga2020role}][\cite{rubin2020psychological}]. 

The Government of India has taken up several measures towards controlling the pandemic, such as strengthening medical care, issuing strict lockdown instructions, carrying out awareness campaigns and so on\footnote{\url{https://www.mygov.in/covid-19/}}. The instructions passed by the government and severity of \textit{Covid-19} has greatly affected the day to day lives of citizens in the country. This is also observed to result in psychological stress among few people in the country. Several measures to reduce this affect are being taken up by both public and private sector organizations in the country. The government has set up helpline centers that aim towards helping people with psychological issues during the pandemic, through telephone. Also, many audio and video awareness clippings are being advertised by many organizations to prevent the psychological stress and to improve awareness among people towards the pandemic. Understanding the mood of people across various parts of the country could help various organizations in taking better measures to help citizens of the country in maintaining better psychological balance. Social Networking platforms such as Twitter, Facebook and so on serve as a source to analyse and understand emotional state of the public. Such platforms play an important role during pandemics, in assessing the mood and mental health of people across the world. Researchers across the globe have made several attempts to understand the emotions of people in various countries towards \textit{Covid-19}, with an aim to help health organizations and government in taking up measures accordingly. However, most of the existing studies summarize the emotions of posts on social networking platforms during a specified time interval, but not on day to day posts. Depicting the emotion of people on a daily basis could help various organizations in understanding the changing mood of people. Hence, we present a web portal- \textit{Mood of India During \textit{Covid-19}}, that is aimed to provide visualizations of  across various states in the country. The tweets posted on twitter related to \textit{Covid-19} are analysed and classified into one of the seven categories that include six emotions - \textit{Anger, Sadness, Happiness, Surprise, Fear and Disgust}[\cite{ekman1992argument}], and \textit{Neutral} category, which are visualized on India Map based on the location from which the tweets have been posted. 


\section{Related Work}
\label{related work}
The emotional state of people plays an important role in responding to a pandemic. Understanding the psychological state of population helps the governments in formulating guidelines and in taking necessary measures that are intended to motivate the public towards taking appropriate measures to prevent spread of pandemics and to restore mental well being amongst vulnerable population. Hence, several studies have been conducted to analyse the emotional state of the public during pandemics, including \textit{Covid-19}.

The influence of various information sources and awareness campaigns aimed towards educating people about the mode of spread, safety measures to be taken against \textit{Covid-19} and so on has been analysed through a open-ended answer based questionnaire based on Situational Awareness Theory and Theory Planned Behaviour [\cite{qazi2020analyzing}]. A user survey has been conducted based on the questionnaire and sentiments of 82 responses received have been analysed. The results of this study indicate that enhanced situational awareness among the people motivates them in adopting better protective measures [\cite{qazi2020analyzing}]. 
Sentiments of 538 responses obtained from an online survey containing open ended questions related to Health Anxiety have been analysed among people in Philippines. The results of this analysis revealed moderate ``\textit{level of symptoms of Hypochondriasis, Attitude on Acquiring \textit{Covid-19}
Avoidance, and Reassurance Seeking}'' among the Philippines population [\cite{nicomedes2020analysis}].

Zhao et al. have analysed the attention of public to events related to \textit{Covid-19} in China and observed an increase in public attention towards information related \textit{Covid-19}. This analysis has been conducted on Chinese social media platform - Sina Microblog, with an aim to help government in formulating better principles in communicating on health related aspects with the public [\cite{zhao2020chinese}]. 
Another study conducted on data from three social media platforms in China during the period of December 1, 2019 to February 15, 2020 aimed to observe public attention, emotion, behavioral response and so on. This data analysis revealed low public attention in the initial stages of the outbreak, delaying the control of \textit{Covid-19}. Also, it has been observed that delayed information broadcasting has triggered negative emotions and resulted in panic buying in many cases [\cite{hou2020assessment}]. Li et al. have analysed more than 17K Weibo posts from 13 January 2020 to 26 January 2020. The analysis was based on identifying the psychological profile of users based on Online Ecological Recognition and predictive machine learning models, and consequently identified the emotions of users. The results indicate a visible decline in life satisfaction and increase in negative emotions [\cite{li2020impact}].

\begin{figure}
  \centering
  \includegraphics[width = \linewidth]{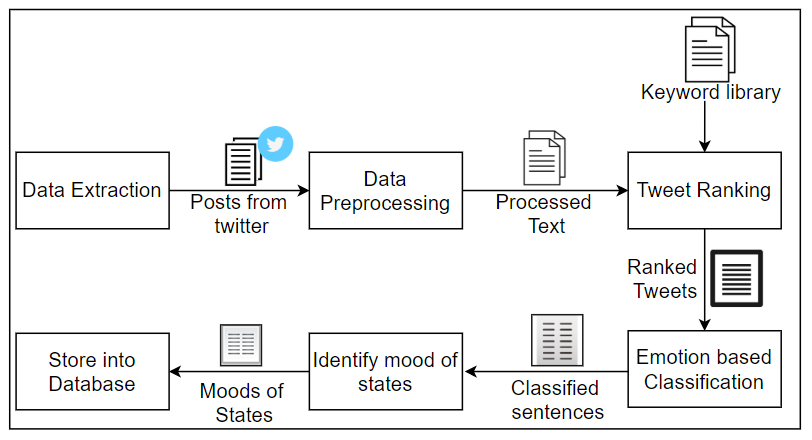}
  \caption{Overview of Approach for \textit{Mood of India During\textit{Covid-19}}}
  \label{fig:app}
\end{figure}
Twitter is one of the most commonly used platform and a rich medium to analyse various factors of population such as public sentiments, public response to a situation across the world, predict outbreaks of diseases and so on. Posts on twitter during 14 January 2020 and 28 January 2020 related \textit{Covid-19} have been extracted have been extracted to understand the changes in sentiments and opinions among people towards \textit{Covid-19}. It has been observed that negative emotion tweets are observed in areas having more number of cases reported [\cite{medford2020infodemic}]. Sharma et al. have designed a dashboard that identifies misinformation being spread with respect to \textit{Covid-19}, reactions of users to various emergency policies, country-wide sentiments and so on, and displays them on the dashboard [\cite{sharma2020coronavirus}]. Considering the importance of posts on twitter, a multilanguage twitter dataset has been created by Lopez et al. This dataset is also expected to provide insights on public response towards the pandemic across several countries and multiple languages [\cite{lopez2020understanding}].
About 364K posts on twitter during 31 December 2019 to 6 February 2020 have been extracted and analysed to predict the outbreak of \textit{Covid-19}. It has been observed that the number of tweets all over the world were directly proportional to the number of cases being reported in respective countries [\cite{jahanbin2020using}]. 
Chen et al. have analysed posts on twitter during 23 March 2020 and 5 April 2020 and respective emotions associated with the posts [\cite{chen2020eyes}]. They observed correlations between the nature of terms used in the posts and respective sentiments of the posts. They have classified the terms into two classes - controversial and non-controversial using LDA topic modelling and the results of analysis indicated that posts with controversial terms exhibit a higher level of negative emotions [\cite{chen2020eyes}]. Public emotions have been analysed based on twitter data of around 20 million posts, tweeted during 28 January 2020 and 9 April 2020 [\cite{lwinglobal}]. The tweets have been classified into four emotion categories - fear, anger, sadness and joy. It has been observed that the sentiments have drifted from fear to anger over the 70 day period [\cite{lwinglobal}]. Barkur et al. have considered about 24K tweets on Twitter during 25 March 2020 and 28 March 2020 to analyse public sentiments on \textit{Covid-19} after the government has issued lockdown instructions [\cite{barkur2020sentiment}]. It has been observed that positive sentiments have outnumbered the negative ones in the country [\cite{barkur2020sentiment}]. 

In spite of several studies being conducted on psychological state of public across the world, there is limited research in understanding the psychological state of public in India. Also, most of the existing studies deal with data only during specific timelines. In addition to this, to the best of knowledge, we are not aware of any real time, streaming twitter datasets, that are specific to India. Hence, we present an interactive web portal, aimed to display the mood of India during \textit{Covid-19}, based on streaming data of twitter. This web portal provides visualizations of number of Covid cases reported and emotional states of various states in the country. Also, the users are facilitated to select date range and state, and to view respective statistics of emotion change as a graph. 
\begin{table}[]
  \centering
  \begin{tabular}{|c|c|}
    \hline
     Emotion&No.of Keywords \\
     \hline
     Anger&355\\
     Disgust&70\\
     Happiness&553\\
     Surprise&95\\
     Fear&195\\
     Sadness&274\\
     \hline
     
  \end{tabular}
  \newline
  \caption{Number of keywords for each emotion in the dataset}
  \label{tab:emolist}
\end{table}

\section{Design Methodology}
\label{design}

\textit{Mood of India During \textit{Covid-19}} thus provides a platform to view the sentiment of people across in each state on each day. It also helps in viewing the trends in emotion change across the country during a specific interval. 

Twitter has been regarded as one of the richest platforms to assess trends, predict several activities, understand emotions and response of people towards various scenarios and so on. It has been widely used in analysing mental health and emotions of people during crisis situations. Hence, analysing emotions posts on twitter during \textit{Covid-19} in India could help in understanding the emotional states of population across the country.

Several studies have reported that texts in twitter posts could be associated with one of the six basic emotions proposed by Ekman [\cite{ekman1992argument}]. Also, it has been observed that several studies aimed to analyse the sentiment of people based on twitter posts during such pandemics and crisis situations, have classified the twitter posts into the six basic emotions [\cite{do2016analyzing}]. Do et al. have classified emotions on twitter during the outbreak of MERS in Korea during 2015, into seven categories [\cite{do2016analyzing}]. We observed the idea of classifying tweets into emotion categories during \textit{Covid-19}, to be similar to that of classifying them during MERS in Korea [\cite{do2016analyzing}]. Hence, we considered a total seven emotions into which each of the tweets was classified, inline with those considered by Do et al. They included the six basic emotions - Anger, Disgust, Fear, Happiness, Sadness and Surprise, and an additional category - Neutral. 

\textit{Mood of India During \textit{Covid-19}} provides emotion of a state in India through a six step mechanism as shown in Figure \ref{fig:app}.

\textbf{\textit{Step 1 - Data Extraction.}} 
Tweets posted from India with six hashtags related to \textit{Covid-19} are downloaded from Twitter. These hashtags included - \textit{CoronaVirus, Covid-19, India Fight Corona, Covid, Lockdown}
All such tweets \textbf{(not more than 10,000 per day)}, posted during a day are downloaded along with their corresponding locations. Based on the location co-ordinates, corresponding states are identified.

\textbf{\textit{Step 2 - Data Preprocessing.}}
The downloaded tweets are preprocessed using NLTK library. This method first tokenizes all the input sentences in the tweets and the parts of speech of each word in tweet in identified. All nouns and prepositions are removed as they do not convey any emotion. 

\textbf{\textit{Step 3 - Sentence Ranking.}}
An existing dataset containing synonyms of keywords corresponding to each of the six emotions - anger, disgust, fear, happiness, sadness and surprise has been downloaded from a github repository\footnote{\url{https://github.com/timjurka/sentiment/blob/master/sentiment/data/emotions.csv.gz}}. The number of keywords for each of the six emotions is presented in Table \ref{tab:emolist}. The resultant tweets after preprocessing are compared to the set of keywords corresponding to each emotion. The tweets are then assigned ranks with respect to each emotion category based on the percentage of keywords of each category present in the tweet.

\textbf{\textit{Step 4 - Emotion based Classification.}}
The ranks of all emotions in each tweet are analysed and the highest ranked emotion in the tweet is considered to be the emotion of the tweet. The tweets are then labelled with their corresponding emotion. The tweets having no rank for any of the six emotions indicate the absence of keywords corresponding to all six emotions. Such tweets are considered to express neutral emotion and hence are labelled as \textit{Neutral}. All the tweets containing facts also fall into the neutral category.

\textbf{\textit{Step 5 - Identify mood of states.}}
The emotions of all the tweets in a day, in each state are analysed. Tweets having similar emotions are grouped together and sum of tweets conveying the same emotion are calculated. The emotion having the largest number of tweets in each state is identified and the states are labelled corresponding to the identified emotion.

\textbf{\textit{Step 6 - Store into Database.}}
The tweets and states with corresponding emotions are stored into the database, to be displayed onto the web portal as required.

Apart from the tweet data, information related to \textit{Covid-19} cases, including the number of new cases reported, number of recovered cases and the number of deaths reported are extracted using Covid19 API\footnote{\url{https://api.covid19india.org/}}. This data is rendered on the portal as a heat map, plotted on the country map.

Apart from choosing to visualize emotion of states in the country, users can also view emotions of any one of the six cities- Mumbai, Chennai, Pune, Hyderabad, Bangalore and Tirupati.

\section{Development}
\label{dev}
The fundamental motivation behind developing the web portal is to provide insights on feelings of people during \textit{Covid-19}, based on twitter information. We have used the Flask framework as a lightweight back-end system and HTML5 for the front-end for the development, which enables us to serve \textit{Mood of India During Covid-19} as a web portal. Using the python scheduler library, we update the previous day's data precisely at 16:00 hours daily. We retrieve and store two kinds of data - \textit{Covid-19} Case data from \textit{covid19India}\footnote{\url{https://api.covid19india.org/}} API and Daily tweet data. We then convert this data into .csv files for processing and generate JSON files with meta-data for reducing the stress on the backend. 

\begin{figure}
  \centering
  \includegraphics[width = \linewidth]{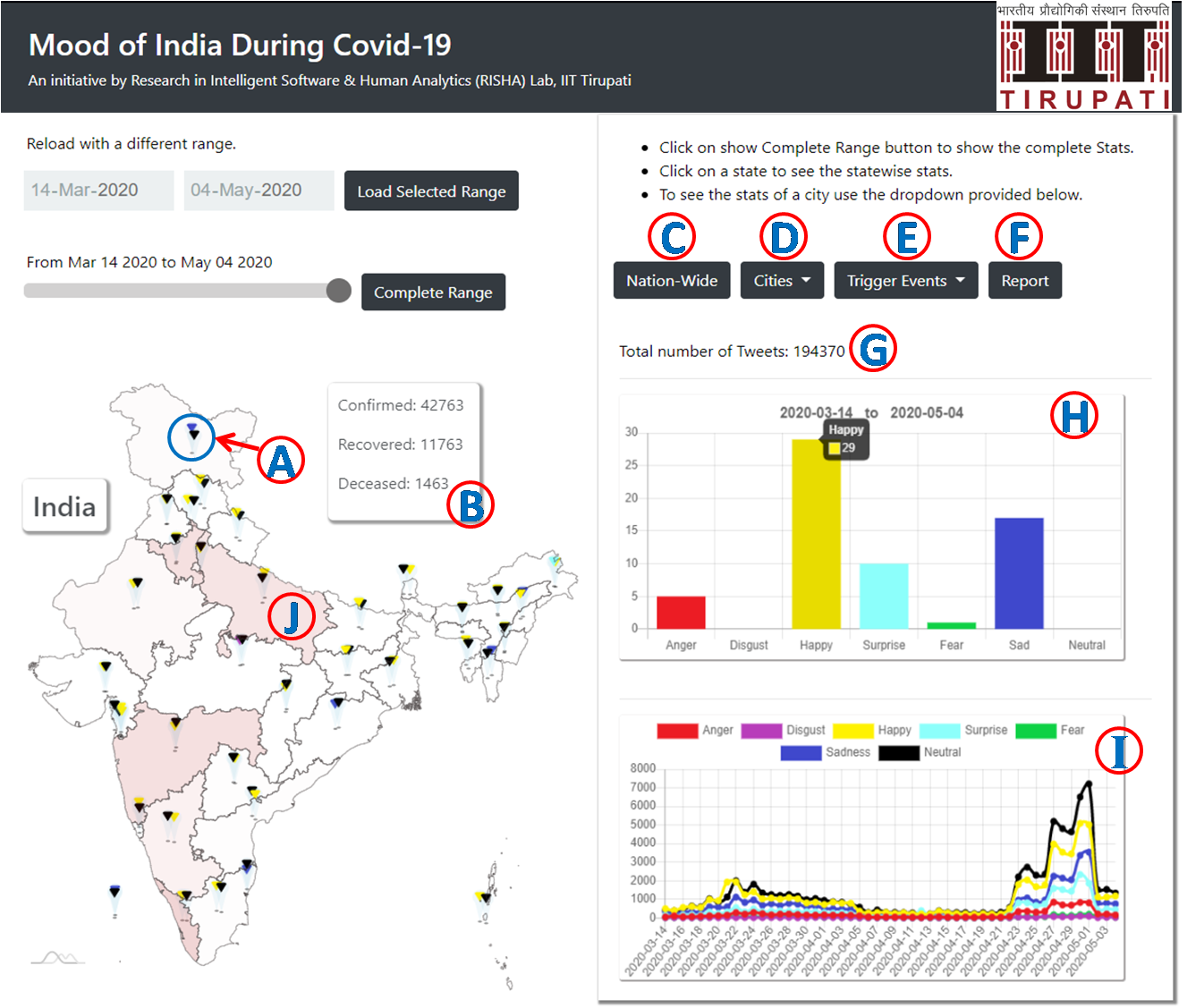}
  \caption{Landing Page of \textit{Mood of India During Covid-19}, depicting [A] Top two emotions of the state, [B] \textit{Covid-19} statistics of the country during the selected range, [C] Nation-Wide button to enable viewing nation-wide statistics, [D] Cities dropdown, to choose desired cities, [E] Trigger Events dropdown to select the desired Trigger Events during the selected date range, [F] Report button to view report of mood of the country during each day in the selected date range, [G] Total number of Tweets analysed to display the graphs,[H] Bar graph displaying emotions during the selected date range, [I] Line graph depicting trends of mood changes during the selected date range and [J] Intensity of \textit{Covid-19} cases.}
  \label{fig:landing}
\end{figure}

Using the tweet data, we generate two kinds of graphs, a bar graph for a single day evaluation and a line graph for evaluation of a continuous range of dates. These graphs are present for India, it's states and cities. \textit{ChartJS} has been used to generate all the graphs in the application. The \textit{Covid-19} case data contains the number of state-wise Confirmed, Recovered and Deceased cases, which is shown to the right of the Map. It is visualized by generating a Heat Map of India. The Map shows state-wise \textit{Covid-19} confirmed cases based on the selected date. We built the Map using an SVG image with all the states listed in it and programmed it manually to respond to user interaction. By combining the slider and the Map, the user can select a specific state and know the statistics at any point in time. 
We plot pins in the graph showing the top two highest emotion/emotions in that state. To convert the geographical co-ordinates into a position on the Map, we use the \textit{amCharts} library. A report is generated for the selected date range, which summarises the available details.

\begin{figure}
  \centering
  \includegraphics[width = \linewidth]{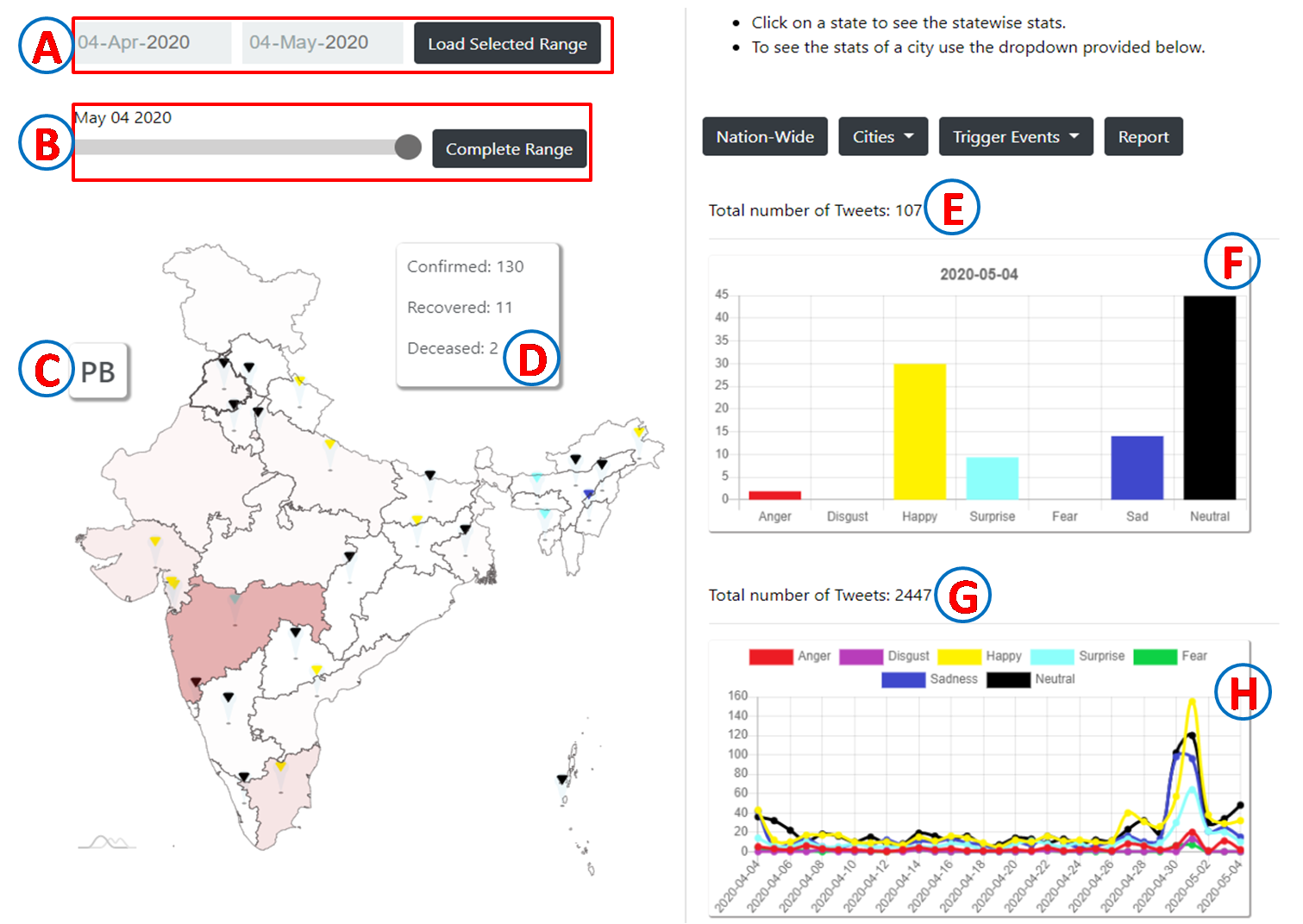}
  \caption{Mood of \textit{Punjab} state depicting [A] Selected Date range, [B] Selected date among the dates in date range on the slider, [C] State Code, [D] \textit{Covid-19} case data on the date selected on slider, [E] Total number of tweets analysed to plot the bar graph on the date selected on slider, [F] Bar graph representing emotions on the date selected on slider in \textit{Punjab}, [G] Total number of tweets analysed to plot the line graph during the selected date range and [H] Line graph representing change in emotions during the selected date range in \textit{Punjab}}
  \label{fig:range}
\end{figure}

\begin{figure}
  \centering
  \includegraphics{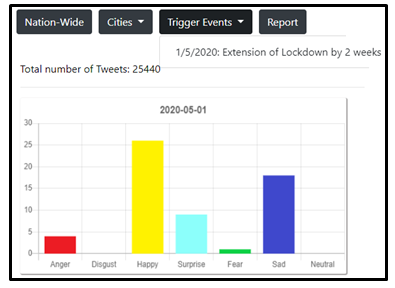}
  \caption{Bar graph representing mood of India on one of the trigger events, May 1, 2020}
  \label{fig:trigger}
\end{figure}

\section{Results}
\label{results}

The web portal, \textit{Mood of India During Covid-19}, displays the emotion of population towards \textit{Covid-19}, across India, with an aim to provide insights about the mood of the country. When the portal is visited on any specific day, emotions of the country from 14 March,2020 to the present day are loaded by default based on twitter data during the range, as shown in Figure \ref{fig:landing}. Two top most emotions among the seven categories are loaded for each state and visualized on the map as shown in [A] of Figure \ref{fig:landing}. A heat map based on the number of new cases reported is displayed on the map. Darker regions on the map indicate higher number of cases being reported, as shown in [J] of Figure \ref{fig:landing}. The emotion based statistics during the range are represented in the bar graph, as shown in [H] of Figure \ref{fig:landing} and the changing trends of emotions during the range are presented b ythe line graph, as shown in [I] of Figure \ref{fig:landing}. Total number of tweets that are analysed to identify the emotions are mentioned as shown in [G] of Figure \ref{fig:landing}. Statistics of any state in the country could be obtained by clicking on the desired state in map. City wide statistics for any of the six cities mentioned in the \textit{Design} section could be obtained by selecting the city from the dropdown presented in [D] of Figure \ref{fig:landing}. Users can also choose to view the mood of population on few specific events, which we termed as \textit{Trigger Events}, by selecting the desired event from the dropdown shown in [E] of Figure \ref{fig:landing}. Nation-wide statistics could be obtained by clicking on \textit{Nation} button, shown in [C] of Figure \ref{fig:landing}. A report of the emotions could be generated based on users' selection of date range, by clicking on the \textit{Report} button, shown in [F] of Figure \ref{fig:landing}. The report includes mood of people in the country, represented as a bar graph, similar to [H] of Figure \ref{fig:landing} and the \textit{Covid-19} statistics of Confirmed, Recovered and Deceased cases for each day in the date range. Users can also select to view the mood of states, including the number of tweets expressing each of the six emotions on each day in the date range along with the statistics of \textit{Covid-19} case status. Reports could thus be useful for policy makers in taking into consideration the mood of the country while introducing policies against \textit{Covid-19}. 

\begin{table}[]
  \fontsize{8}{12}\selectfont
  
  \begin{tabular}{|c|c|c|c|c|c|c|c|c|c|c|c|}
  \hline
\backslashbox{\textit{\textbf{State}}}{\textit{\textbf{Emotions}}}&\textit{\textbf{A}}&\textit{\textbf{D}}&\textit{\textbf{H}}& \textit{\textbf{S}}&\textit{\textbf{F}}& \textit{\textbf{SA}}&\textit{\textbf{N}}&\textit{\textbf{Tot}}&\textit{\textbf{Con}}&\textit{\textbf{Rec}}&\textit{\textbf{Dec}}\\

\hline

Bihar & 169 & 32 & 1491 & 464 & 48 & 1011 & 1580 & 4795& 528 & 127 & 4\\
\hline
Goa & 469 & 52 & 2955 & 1017 & 97 & 2154 & 3355 &10099& 7 & 7 & 0
\\
\hline
Punjab & 857 & 122 & 4825 & 1680 & 253 & 3379 & 5136 &16252& 1232 & 128 & 23
\\
\hline
Uttar& 392 & 14 & 2344 & 607 & 45 & 1143 & 2551 & 7096 & 2766 & 802 & 50
\\
Pradesh&&&&&&&&&&&\\
\hline
Chandigarh & 1740 & 177 & 8826 & 3439 & 406 & 6029 & 12380& 32997 & 116 & 21 & 1
\\
\hline
Dadra and & 90 & 0 & 598 & 170 & 28 & 354 & 570& 1810 & 0 & 0 & 0
\\
Nagar Haveli & & & & & & &  & & & &\\
\hline
Daman and Diu & 152 & 8 & 1002 & 297 & 37 & 627 & 1042& 3165 & 0 & 0 & 0
\\
\hline
Delhi & 10284 & 1087 & 58617 & 22053 & 2450 & 37439 & 69640& 201570 & 4898 & 1431 & 64
\\
\hline
Rajasthan & 199 & 12 & 1150 & 459 & 53 & 606 & 1472& 3951 & 3061 & 1438 & 77
\\
\hline
Tamil Nadu & 785 & 89 & 4826 & 2084 & 213 & 2904 & 5956& 16857 & 3550 & 1409 & 31
\\
\hline
West Bengal & 459 & 37 & 2621 & 931 & 124 & 1624 & 3219 & 9015 & 1259 & 218 & 133
\\
\hline
Maharashtra & 377 & 41 & 2209 & 882 & 103 & 1300 & 2797& 7709 & 14541 & 2465 & 582
\\
\hline
Telangana & 974 & 118 & 4673 & 1663 & 194 & 2443 & 7004& 17069 & 1085 & 585 & 29
\\
\hline
Pondicherry & 154 & 25 & 1024 & 421 & 39 & 632 & 1507& 3802 & 9 & 6 & 0
\\
\hline
Andhra& 7 & 2 & 124 & 27 & 1 & 40 & 91& 292 & 1650 & 524 & 33
\\
Pradesh&&&&&&&&&&&\\
\hline
Chhattisgarh & 80 & 2 & 450 & 142 & 20 & 287 & 609& 1590 & 44 & 36 & 0
\\
\hline
Gujarat & 36 & 11 & 318 & 99 & 14 & 218 & 418& 1114 & 5804 & 1195 & 319
\\
\hline
Karnataka & 25 & 0 & 100 & 21 & 22 & 44 & 122& 334 & 651 & 321 & 27
\\
\hline
Manipur & 6 & 6 & 110 & 23 & 2 & 119 & 155 & 421 & 2 & 2 & 0
\\
\hline
Haryana & 304 & 10 & 2056 & 642 & 88 & 995 & 2584& 6679 & 517 & 253 & 6
\\
\hline
Jharkhand & 46 & 2 & 439 & 114 & 27 & 373 & 489 & 1490  & 115 & 27 & 3
\\
\hline
Andaman and  & 41 & 0 & 151 & 53 & 2 & 86 & 180& 513 & 33 & 32 & 0
\\
Nicobar Islands&&&&&&&&&&&
\\
\hline

Tripura & 36 & 12 & 409 & 129 & 3 & 168 & 698& 1455 & 29 & 2 & 0
\\
\hline
Orissa & 0 & 0 & 7 & 4 & 0 & 7 & 18& 36 & 169 & 60 & 1
\\
\hline
Meghalaya & 4 & 0 & 86 & 47 & 12 & 47 & 83& 279 & 12 & 10 & 1
\\
\hline
Sikkim & 5 & 0 & 9 & 5 & 0 & 6 & 30 & 55 & 0 & 0 & 0
\\
\hline
Arunachal& 0 & 0 & 12 & 8 & 0 & 0 & 6 & 26 & 1 & 1 & 0
\\
Pradesh&&&&&&&&&&&\\
\hline
Himachal& 0 & 0 & 11 & 0 & 0 & 1 & 0 & 12 & 41 & 35 & 2
\\
Pradesh&&&&&&&&&&&\\
\hline
Nagaland & 3 & 0 & 67 & 20 & 0 & 41 & 48 & 179 & 0 & 0 & 0
\\
\hline
Madhya& 0 & 2 & 1 & 0 & 0 & 0 & 5 & 8 & 2942 & 856 & 166
\\
Pradesh&&&&&&&&&&&\\
\hline
Mizoram & 0 & 0 & 0 & 0 & 0 & 0 & 0 & 0 & 1 & 0 & 0
\\
\hline
Kerala & 48 & 1 & 462 & 199 & 39 & 294 & 745& 1788 & 500 & 462 & 4
\\
\hline
Assam & 9 & 4 & 208 & 49 & 5 & 177 & 279 & 731 & 43 & 33 & 1
\\
\hline

  \end{tabular}
  \newline
  \caption{Emotions expressed by Tweets in each state and union territory of India and their corresponding \textit{Covid-19} Statistics. The emotions are represented by \textbf{A} - Anger, \textbf{D} - Disgust, \textbf{H} - Happiness, \textbf{S}- Surprise, \textbf{F}- Fear, \textbf{SA} - Sadness \textbf{N} - Neutral and \textbf{Tot} - Total number of tweets, with values in terms of number of tweets, and \textit{Covid-19} statistics are represented by \textbf{Con} - Confirmed, \textbf{Rec} - Recovered and \textbf{Dec} - Deceased, with values in terms of number of cases.  }
  \label{tab:stats}
\end{table}

Table \ref{tab:stats} depicts the number of tweets classified into each of the seven categories, for every state and union territory of India, along with \textit{Covid-19} statistics in the corresponding regions during March 14, 2020 to May 6, 2020. 

Figure \ref{fig:range} displays the emotions of the population during the selected date range(04-Apr-2020 to 04-May-2020), as shown in [A] of Figure \ref{fig:range}. 
When a state, \textit{Punjab}, is clicked on the map, emotions of \textit{Punjab} on the day selected in the slider (May 04, 2020, as represented by [B] of Figure \ref{fig:range}), in the specified range are displayed on the bar graph, as shown in [F] of Figure \ref{fig:range}. [C] of Figure \ref{fig:range} indicates the code of the state(PB) and [D] of Figure \ref{fig:range} indicates \textit{Covid-19} statistics in Punjab, on May 4, 2020. The total number of \textit{Covid-19} related tweets used for analysis on May 4, 2020 are represented by [E] of Figure \ref{fig:range}. It could be observed from the bar graph that around 30\% of the tweets expressed \textit{Happiness}, while 45\% of the tweets expressed \textit{Neutral} emotion on May 4,2020. The line graph, shown in [H] of Figure \ref{fig:range} indicates the change in emotion during the specified range in the selected state. [G] of Figure \ref{fig:range} indicates the total number of tweets from \textit{Punjab} considered during the range for evaluation. It could be observed that tweets expressing \textit{Happiness} have reduced during the range May 1, 2020 and May 4, 2020. The increasing number of deceased cases in Punjab could explain the changing mood of people.

The bar graph in Figure \ref{fig:trigger} indicates the distribution of emotions among the people of State, when the Extension of Lockdown by 2 weeks has been announced on May 1 2020. This graph is generated by selecting the Trigger Event - \textit{Extension of Lockdown by 2 weeks}. The graph displays mixed emotions, with about 26\% of the tweets expressing happiness, 18\% expressing sadness and so on, indicating that most of the people in the state are happy about the announcement.

\section{Discussion and Limitations}
\textit{Mood of India During \textit{Covid-19}} has been developed as a webportal, that is aimed to provide information about emotions of the population during the pandemic. Data in the form of posts from twitter are mined on a daily basis, along with the location from which they are posted. Each of tweets is classified into one of the seven categories, based on the presence of percentage of keywords belonging to the six emotion categories. The tweets containing facts are classified into the neutral category.

Currently, our database consists of about 86K tweets, during the period of March and April. However, more number of tweets for the two months could be mined gradually. Also, the maximum number of tweets being mined per day currently is 5K tweets from each state. Though more number of tweets could be mined, we observe that the total number of \textit{Covid-19} related posts are around 10k-12K per day across the country, on an average. Hence, we assume that the current number of tweets would suffice for analysis. However, this number could be increased if the number of tweets are observed to raise. 

Tweets are being classified into respective emotions based on comparison of the processed tweets with set of keywords present in the dataset extracted from the github repository\footnote{\url{https://github.com/timjurka/sentiment/blob/master/sentiment/data/emotions.csv.gz}}. Tweets containing words that would belong to an emotion, but not present in the set of respective keywords would not be identified to express the specific emotion. The existing list of keywords contains a minimum of 70 keywords per emotion. Since the number of keywords in the dataset for each emotion are not equal, the classification might be inclined towards emotions with more number of keywords in the dataset. Identifying and implementing Natural Language Processing and Machine Learning techniques that result in better accuracies could be used to improve the classification of tweets and consequently in better predicting the mood of the country.

The existing portal displays mood of the population only upto state level. Mood of population in only six cities - Pune, Hyderabad, Mumbai, Tirupati, Chennai and Bangalore is represented. Representing the emotions at multiple levels such as nation, state, city, district and so on, could help in understanding the emotional states of people belonging to more specific regions.

\section{Conclusion and Future Work}
Considering the importance of understanding public emotions and the affects on psychological state of people during a crisis, in this paper, we present a web portal to identify the mood of India during the current \textit{Covid-19} pandemic. The web portal marks top most emotions of people across various states and cities in the country based on the sentiments associated with tweets in each region. The emotions of tweets are identified by ranking tweets based on comparison of words in the tweets with pre-loaded list of keywords for each emotion. The results of emotions in a specific state in the country are displayed on the country map on any day selected by user from March 14, 2020. Also, a heat map of number of new cases reported in the country across all the states is plotted on the country map. The trends of emotions changing across the country in a selected date range are also visualized as line graphs. Users can select to view information related to mood of the country. Also, bar graphs depicting percentage of tweets expressing specific emotions are presented to the users. Viewing the country map on any specific day could help users in understanding the emotion of the region with respect to number of \textit{Covid-19} cases reported in the region.

In the future versions, we plan to increase the number of tweets being considered for emotion analysis. Also, we plan to mine more number of tweets related to \textit{Covid-19} in the country during the two month period of March and April to increase the accuracy of emotions being displayed. We also plan to improve the accuracy of classification model by exploring newer NLP and ML based approaches that could classify the tweets based on emotions. Furthermore, the existing portal could be improved to display mood of population with more specificity, which could include districts and cities of each state.


\bibliographystyle{apacite}
\bibliography{interactapasample}

\begin{thebibliography}{}

\bibitem [\protect \citeauthoryear {%
Barkur%
\ \BBA {} Vibha%
}{%
Barkur%
\ \BBA {} Vibha%
}{%
{\protect \APACyear {2020}}%
}]{%
barkur2020sentiment}
\APACinsertmetastar {%
barkur2020sentiment}%
\begin{APACrefauthors}%
Barkur, G.%
\BCBT {}\ \BBA {} Vibha, G\BPBI B\BPBI K.%
\end{APACrefauthors}%
\unskip\
\newblock
\APACrefYearMonthDay{2020}{}{}.
\newblock
{\BBOQ}\APACrefatitle {Sentiment Analysis of Nationwide Lockdown due to COVID
  19 Outbreak: Evidence from India} {Sentiment analysis of nationwide lockdown
  due to covid 19 outbreak: Evidence from india}.{\BBCQ}
\newblock
\APACjournalVolNumPages{Asian Journal of Psychiatry}{}{}{}.
\PrintBackRefs{\CurrentBib}

\bibitem [\protect \citeauthoryear {%
CDC%
}{%
CDC%
}{%
{\protect \APACyear {2020}}%
}]{%
cdccovid}
\APACinsertmetastar {%
cdccovid}%
\begin{APACrefauthors}%
CDC.%
\end{APACrefauthors}%
\unskip\
\newblock
\APACrefYearMonthDay{2020}{}{}.
\newblock
{\BBOQ}\APACrefatitle {Stress and Coping - Coronavirus Disease 2019 (COVID-19)}
  {Stress and coping - coronavirus disease 2019 (covid-19)}.{\BBCQ}
\newblock

\PrintBackRefs{\CurrentBib}

\bibitem [\protect \citeauthoryear {%
Chen%
, Lyu%
, Yang%
, Wang%
\BCBL {}\ \BBA {} Luo%
}{%
Chen%
\ \protect \BOthers {.}}{%
{\protect \APACyear {2020}}%
}]{%
chen2020eyes}
\APACinsertmetastar {%
chen2020eyes}%
\begin{APACrefauthors}%
Chen, L.%
, Lyu, H.%
, Yang, T.%
, Wang, Y.%
\BCBL {}\ \BBA {} Luo, J.%
\end{APACrefauthors}%
\unskip\
\newblock
\APACrefYearMonthDay{2020}{}{}.
\newblock
{\BBOQ}\APACrefatitle {In the Eyes of the Beholder: Sentiment and Topic
  Analyses on Social Media Use of Neutral and Controversial Terms for COVID-19}
  {In the eyes of the beholder: Sentiment and topic analyses on social media
  use of neutral and controversial terms for covid-19}.{\BBCQ}
\newblock
\APACjournalVolNumPages{arXiv preprint arXiv:2004.10225}{}{}{}.
\PrintBackRefs{\CurrentBib}

\bibitem [\protect \citeauthoryear {%
Do%
, Lim%
, Kim%
\BCBL {}\ \BBA {} Choi%
}{%
Do%
\ \protect \BOthers {.}}{%
{\protect \APACyear {2016}}%
}]{%
do2016analyzing}
\APACinsertmetastar {%
do2016analyzing}%
\begin{APACrefauthors}%
Do, H\BPBI J.%
, Lim, C\BHBI G.%
, Kim, Y\BPBI J.%
\BCBL {}\ \BBA {} Choi, H\BHBI J.%
\end{APACrefauthors}%
\unskip\
\newblock
\APACrefYearMonthDay{2016}{}{}.
\newblock
{\BBOQ}\APACrefatitle {Analyzing emotions in twitter during a crisis: A case
  study of the 2015 Middle East Respiratory Syndrome outbreak in Korea}
  {Analyzing emotions in twitter during a crisis: A case study of the 2015
  middle east respiratory syndrome outbreak in korea}.{\BBCQ}
\newblock
\BIn{} \APACrefbtitle {2016 international conference on big data and smart
  computing (BigComp)} {2016 international conference on big data and smart
  computing (bigcomp)}\ (\BPGS\ 415--418).
\PrintBackRefs{\CurrentBib}

\bibitem [\protect \citeauthoryear {%
Douglas%
, Douglas%
, Harrigan%
\BCBL {}\ \BBA {} Douglas%
}{%
Douglas%
\ \protect \BOthers {.}}{%
{\protect \APACyear {2009}}%
}]{%
douglas2009preparing}
\APACinsertmetastar {%
douglas2009preparing}%
\begin{APACrefauthors}%
Douglas, P\BPBI K.%
, Douglas, D\BPBI B.%
, Harrigan, D\BPBI C.%
\BCBL {}\ \BBA {} Douglas, K\BPBI M.%
\end{APACrefauthors}%
\unskip\
\newblock
\APACrefYearMonthDay{2009}{}{}.
\newblock
{\BBOQ}\APACrefatitle {Preparing for pandemic influenza and its aftermath:
  mental health issues considered} {Preparing for pandemic influenza and its
  aftermath: mental health issues considered}.{\BBCQ}
\newblock
\APACjournalVolNumPages{International journal of emergency mental
  health}{11}{3}{137}.
\PrintBackRefs{\CurrentBib}

\bibitem [\protect \citeauthoryear {%
Ekman%
}{%
Ekman%
}{%
{\protect \APACyear {1992}}%
}]{%
ekman1992argument}
\APACinsertmetastar {%
ekman1992argument}%
\begin{APACrefauthors}%
Ekman, P.%
\end{APACrefauthors}%
\unskip\
\newblock
\APACrefYearMonthDay{1992}{}{}.
\newblock
{\BBOQ}\APACrefatitle {An argument for basic emotions} {An argument for basic
  emotions}.{\BBCQ}
\newblock
\APACjournalVolNumPages{Cognition \& emotion}{6}{3-4}{169--200}.
\PrintBackRefs{\CurrentBib}

\bibitem [\protect \citeauthoryear {%
Garcia%
}{%
Garcia%
}{%
{\protect \APACyear {2017}}%
}]{%
garcia2017neurobiology}
\APACinsertmetastar {%
garcia2017neurobiology}%
\begin{APACrefauthors}%
Garcia, R.%
\end{APACrefauthors}%
\unskip\
\newblock
\APACrefYearMonthDay{2017}{}{}.
\newblock
{\BBOQ}\APACrefatitle {Neurobiology of fear and specific phobias} {Neurobiology
  of fear and specific phobias}.{\BBCQ}
\newblock
\APACjournalVolNumPages{Learning \& Memory}{24}{9}{462--471}.
\PrintBackRefs{\CurrentBib}

\bibitem [\protect \citeauthoryear {%
Garriga%
\ \protect \BOthers {.}}{%
Garriga%
\ \protect \BOthers {.}}{%
{\protect \APACyear {2020}}%
}]{%
garriga2020role}
\APACinsertmetastar {%
garriga2020role}%
\begin{APACrefauthors}%
Garriga, M.%
, Agasi, I.%
, Fedida, E.%
, Pinz{\'o}n-Espinosa, J.%
, Vazquez, M.%
, Pacchiarotti, I.%
\BCBL {}\ \BBA {} Vieta, E.%
\end{APACrefauthors}%
\unskip\
\newblock
\APACrefYearMonthDay{2020}{}{}.
\newblock
{\BBOQ}\APACrefatitle {The role of Mental Health Home Hospitalization Care
  during the COVID-19 pandemic} {The role of mental health home hospitalization
  care during the covid-19 pandemic}.{\BBCQ}
\newblock
\APACjournalVolNumPages{Acta Psychiatrica Scandinavica}{}{}{}.
\PrintBackRefs{\CurrentBib}

\bibitem [\protect \citeauthoryear {%
Hou%
, Du%
, Jiang%
, Zhou%
\BCBL {}\ \BBA {} Lin%
}{%
Hou%
\ \protect \BOthers {.}}{%
{\protect \APACyear {2020}}%
}]{%
hou2020assessment}
\APACinsertmetastar {%
hou2020assessment}%
\begin{APACrefauthors}%
Hou, Z.%
, Du, F.%
, Jiang, H.%
, Zhou, X.%
\BCBL {}\ \BBA {} Lin, L.%
\end{APACrefauthors}%
\unskip\
\newblock
\APACrefYearMonthDay{2020}{}{}.
\newblock
{\BBOQ}\APACrefatitle {Assessment of public attention, risk perception,
  emotional and behavioural responses to the COVID-19 outbreak: social media
  surveillance in China} {Assessment of public attention, risk perception,
  emotional and behavioural responses to the covid-19 outbreak: social media
  surveillance in china}.{\BBCQ}
\newblock
\APACjournalVolNumPages{Risk Perception, Emotional and Behavioural Responses to
  the COVID-19 Outbreak: Social Media Surveillance in China (3/6/2020)}{}{}{}.
\PrintBackRefs{\CurrentBib}

\bibitem [\protect \citeauthoryear {%
Jahanbin%
\ \BBA {} Rahmanian%
}{%
Jahanbin%
\ \BBA {} Rahmanian%
}{%
{\protect \APACyear {2020}}%
}]{%
jahanbin2020using}
\APACinsertmetastar {%
jahanbin2020using}%
\begin{APACrefauthors}%
Jahanbin, K.%
\BCBT {}\ \BBA {} Rahmanian, V.%
\end{APACrefauthors}%
\unskip\
\newblock
\APACrefYearMonthDay{2020}{}{}.
\newblock
{\BBOQ}\APACrefatitle {Using twitter and web news mining to predict COVID-19
  outbreak} {Using twitter and web news mining to predict covid-19
  outbreak}.{\BBCQ}
\newblock
\APACjournalVolNumPages{Asian Pacific Journal of Tropical Medicine}{}{}{13}.
\PrintBackRefs{\CurrentBib}

\bibitem [\protect \citeauthoryear {%
J\BHBI B.~Li%
\ \protect \BOthers {.}}{%
J\BHBI B.~Li%
\ \protect \BOthers {.}}{%
{\protect \APACyear {2020}}%
}]{%
li2020chinese}
\APACinsertmetastar {%
li2020chinese}%
\begin{APACrefauthors}%
Li, J\BHBI B.%
, Yang, A.%
, Dou, K.%
, Wang, L\BHBI X.%
, Zhang, M\BHBI C.%
\BCBL {}\ \BBA {} Lin, X.%
\end{APACrefauthors}%
\unskip\
\newblock
\APACrefYearMonthDay{2020}{}{}.
\newblock
{\BBOQ}\APACrefatitle {Chinese public’s knowledge, perceived severity, and
  perceived controllability of the COVID-19 and their associations with
  emotional and behavioural reactions, social participation, and precautionary
  behaviour: A national survey} {Chinese public’s knowledge, perceived
  severity, and perceived controllability of the covid-19 and their
  associations with emotional and behavioural reactions, social participation,
  and precautionary behaviour: A national survey}.{\BBCQ}
\newblock

\PrintBackRefs{\CurrentBib}

\bibitem [\protect \citeauthoryear {%
S.~Li%
, Wang%
, Xue%
, Zhao%
\BCBL {}\ \BBA {} Zhu%
}{%
S.~Li%
\ \protect \BOthers {.}}{%
{\protect \APACyear {2020}}%
}]{%
li2020impact}
\APACinsertmetastar {%
li2020impact}%
\begin{APACrefauthors}%
Li, S.%
, Wang, Y.%
, Xue, J.%
, Zhao, N.%
\BCBL {}\ \BBA {} Zhu, T.%
\end{APACrefauthors}%
\unskip\
\newblock
\APACrefYearMonthDay{2020}{}{}.
\newblock
{\BBOQ}\APACrefatitle {The impact of COVID-19 epidemic declaration on
  psychological consequences: a study on active Weibo users} {The impact of
  covid-19 epidemic declaration on psychological consequences: a study on
  active weibo users}.{\BBCQ}
\newblock
\APACjournalVolNumPages{International journal of environmental research and
  public health}{17}{6}{2032}.
\PrintBackRefs{\CurrentBib}

\bibitem [\protect \citeauthoryear {%
Lin%
, Savoia%
, Agboola%
\BCBL {}\ \BBA {} Viswanath%
}{%
Lin%
\ \protect \BOthers {.}}{%
{\protect \APACyear {2014}}%
}]{%
lin2014have}
\APACinsertmetastar {%
lin2014have}%
\begin{APACrefauthors}%
Lin, L.%
, Savoia, E.%
, Agboola, F.%
\BCBL {}\ \BBA {} Viswanath, K.%
\end{APACrefauthors}%
\unskip\
\newblock
\APACrefYearMonthDay{2014}{}{}.
\newblock
{\BBOQ}\APACrefatitle {What have we learned about communication inequalities
  during the H1N1 pandemic: a systematic review of the literature} {What have
  we learned about communication inequalities during the h1n1 pandemic: a
  systematic review of the literature}.{\BBCQ}
\newblock
\APACjournalVolNumPages{BMC Public Health}{14}{1}{484}.
\PrintBackRefs{\CurrentBib}

\bibitem [\protect \citeauthoryear {%
Lopez%
, Vasu%
\BCBL {}\ \BBA {} Gallemore%
}{%
Lopez%
\ \protect \BOthers {.}}{%
{\protect \APACyear {2020}}%
}]{%
lopez2020understanding}
\APACinsertmetastar {%
lopez2020understanding}%
\begin{APACrefauthors}%
Lopez, C\BPBI E.%
, Vasu, M.%
\BCBL {}\ \BBA {} Gallemore, C.%
\end{APACrefauthors}%
\unskip\
\newblock
\APACrefYearMonthDay{2020}{}{}.
\newblock
{\BBOQ}\APACrefatitle {Understanding the perception of COVID-19 policies by
  mining a multilanguage Twitter dataset} {Understanding the perception of
  covid-19 policies by mining a multilanguage twitter dataset}.{\BBCQ}
\newblock
\APACjournalVolNumPages{arXiv preprint arXiv:2003.10359}{}{}{}.
\PrintBackRefs{\CurrentBib}

\bibitem [\protect \citeauthoryear {%
Lwin%
\ \protect \BOthers {.}}{%
Lwin%
\ \protect \BOthers {.}}{%
{\protect \APACyear {{\protect \bibnodate {}}}}%
}]{%
lwinglobal}
\APACinsertmetastar {%
lwinglobal}%
\begin{APACrefauthors}%
Lwin, M.%
, Sheldenkar, A.%
, Lu, J.%
, Schulz, P.%
, Shin, W.%
, Gupta, R.%
\BCBL {}\ \BBA {} Yang, Y.%
\end{APACrefauthors}%
\unskip\
\newblock
\APACrefYearMonthDay{{\protect \bibnodate {}}}{}{}.
\newblock
{\BBOQ}\APACrefatitle {Global sentiments surrounding the COVID-19 pandemic on
  Twitter} {Global sentiments surrounding the covid-19 pandemic on
  twitter}.{\BBCQ}
\newblock

\PrintBackRefs{\CurrentBib}

\bibitem [\protect \citeauthoryear {%
Mamun%
\ \BBA {} Griffiths%
}{%
Mamun%
\ \BBA {} Griffiths%
}{%
{\protect \APACyear {2020}}%
}]{%
mamun2020first}
\APACinsertmetastar {%
mamun2020first}%
\begin{APACrefauthors}%
Mamun, M\BPBI A.%
\BCBT {}\ \BBA {} Griffiths, M\BPBI D.%
\end{APACrefauthors}%
\unskip\
\newblock
\APACrefYearMonthDay{2020}{}{}.
\newblock
{\BBOQ}\APACrefatitle {First COVID-19 suicide case in Bangladesh due to fear of
  COVID-19 and xenophobia: possible suicide prevention strategies} {First
  covid-19 suicide case in bangladesh due to fear of covid-19 and xenophobia:
  possible suicide prevention strategies}.{\BBCQ}
\newblock
\APACjournalVolNumPages{Asian journal of psychiatry}{51}{}{102073}.
\PrintBackRefs{\CurrentBib}

\bibitem [\protect \citeauthoryear {%
Medford%
, Saleh%
, Sumarsono%
, Perl%
\BCBL {}\ \BBA {} Lehmann%
}{%
Medford%
\ \protect \BOthers {.}}{%
{\protect \APACyear {2020}}%
}]{%
medford2020infodemic}
\APACinsertmetastar {%
medford2020infodemic}%
\begin{APACrefauthors}%
Medford, R\BPBI J.%
, Saleh, S\BPBI N.%
, Sumarsono, A.%
, Perl, T\BPBI M.%
\BCBL {}\ \BBA {} Lehmann, C\BPBI U.%
\end{APACrefauthors}%
\unskip\
\newblock
\APACrefYearMonthDay{2020}{}{}.
\newblock
{\BBOQ}\APACrefatitle {An" Infodemic": Leveraging High-Volume Twitter Data to
  Understand Public Sentiment for the COVID-19 Outbreak} {An" infodemic":
  Leveraging high-volume twitter data to understand public sentiment for the
  covid-19 outbreak}.{\BBCQ}
\newblock
\APACjournalVolNumPages{medRxiv}{}{}{}.
\PrintBackRefs{\CurrentBib}

\bibitem [\protect \citeauthoryear {%
Nicomedes%
\ \BBA {} Avila%
}{%
Nicomedes%
\ \BBA {} Avila%
}{%
{\protect \APACyear {2020}}%
}]{%
nicomedes2020analysis}
\APACinsertmetastar {%
nicomedes2020analysis}%
\begin{APACrefauthors}%
Nicomedes, C.%
\BCBT {}\ \BBA {} Avila, R.%
\end{APACrefauthors}%
\unskip\
\newblock
\APACrefYearMonthDay{2020}{}{}.
\newblock
{\BBOQ}\APACrefatitle {An Analysis on the Panic of Filipinos During COVID-19
  Pandemic in the Philippines} {An analysis on the panic of filipinos during
  covid-19 pandemic in the philippines}.{\BBCQ}
\newblock

\PrintBackRefs{\CurrentBib}

\bibitem [\protect \citeauthoryear {%
Qazi%
\ \protect \BOthers {.}}{%
Qazi%
\ \protect \BOthers {.}}{%
{\protect \APACyear {2020}}%
}]{%
qazi2020analyzing}
\APACinsertmetastar {%
qazi2020analyzing}%
\begin{APACrefauthors}%
Qazi, A.%
, Qazi, J.%
, Naseer, K.%
, Zeeshan, M.%
, Hardaker, G.%
, Maitama, J\BPBI Z.%
\BCBL {}\ \BBA {} Haruna, K.%
\end{APACrefauthors}%
\unskip\
\newblock
\APACrefYearMonthDay{2020}{}{}.
\newblock
{\BBOQ}\APACrefatitle {Analyzing Situational Awareness through Public Opinion
  to Predict Adoption of Social Distancing Amid Pandemic COVID-19} {Analyzing
  situational awareness through public opinion to predict adoption of social
  distancing amid pandemic covid-19}.{\BBCQ}
\newblock
\APACjournalVolNumPages{Journal of Medical Virology}{}{}{}.
\PrintBackRefs{\CurrentBib}

\bibitem [\protect \citeauthoryear {%
Rubin%
\ \BBA {} Wessely%
}{%
Rubin%
\ \BBA {} Wessely%
}{%
{\protect \APACyear {2020}}%
}]{%
rubin2020psychological}
\APACinsertmetastar {%
rubin2020psychological}%
\begin{APACrefauthors}%
Rubin, G\BPBI J.%
\BCBT {}\ \BBA {} Wessely, S.%
\end{APACrefauthors}%
\unskip\
\newblock
\APACrefYearMonthDay{2020}{}{}.
\newblock
{\BBOQ}\APACrefatitle {The psychological effects of quarantining a city} {The
  psychological effects of quarantining a city}.{\BBCQ}
\newblock
\APACjournalVolNumPages{Bmj}{368}{}{}.
\PrintBackRefs{\CurrentBib}

\bibitem [\protect \citeauthoryear {%
Sharma%
\ \protect \BOthers {.}}{%
Sharma%
\ \protect \BOthers {.}}{%
{\protect \APACyear {2020}}%
}]{%
sharma2020coronavirus}
\APACinsertmetastar {%
sharma2020coronavirus}%
\begin{APACrefauthors}%
Sharma, K.%
, Seo, S.%
, Meng, C.%
, Rambhatla, S.%
, Dua, A.%
\BCBL {}\ \BBA {} Liu, Y.%
\end{APACrefauthors}%
\unskip\
\newblock
\APACrefYearMonthDay{2020}{}{}.
\newblock
{\BBOQ}\APACrefatitle {Coronavirus on social media: Analyzing misinformation in
  Twitter conversations} {Coronavirus on social media: Analyzing misinformation
  in twitter conversations}.{\BBCQ}
\newblock
\APACjournalVolNumPages{arXiv preprint arXiv:2003.12309}{}{}{}.
\PrintBackRefs{\CurrentBib}

\bibitem [\protect \citeauthoryear {%
Shigemura%
, Ursano%
, Morganstein%
, Kurosawa%
\BCBL {}\ \BBA {} Benedek%
}{%
Shigemura%
\ \protect \BOthers {.}}{%
{\protect \APACyear {2020}}%
}]{%
shigemura2020public}
\APACinsertmetastar {%
shigemura2020public}%
\begin{APACrefauthors}%
Shigemura, J.%
, Ursano, R\BPBI J.%
, Morganstein, J\BPBI C.%
, Kurosawa, M.%
\BCBL {}\ \BBA {} Benedek, D\BPBI M.%
\end{APACrefauthors}%
\unskip\
\newblock
\APACrefYearMonthDay{2020}{}{}.
\newblock
{\BBOQ}\APACrefatitle {Public responses to the novel 2019 coronavirus
  (2019-nCoV) in Japan: mental health consequences and target populations}
  {Public responses to the novel 2019 coronavirus (2019-ncov) in japan: mental
  health consequences and target populations}.{\BBCQ}
\newblock
\APACjournalVolNumPages{Psychiatry and clinical neurosciences}{74}{4}{281}.
\PrintBackRefs{\CurrentBib}

\bibitem [\protect \citeauthoryear {%
Shin%
\ \BBA {} Liberzon%
}{%
Shin%
\ \BBA {} Liberzon%
}{%
{\protect \APACyear {2010}}%
}]{%
shin2010neurocircuitry}
\APACinsertmetastar {%
shin2010neurocircuitry}%
\begin{APACrefauthors}%
Shin, L\BPBI M.%
\BCBT {}\ \BBA {} Liberzon, I.%
\end{APACrefauthors}%
\unskip\
\newblock
\APACrefYearMonthDay{2010}{}{}.
\newblock
{\BBOQ}\APACrefatitle {The neurocircuitry of fear, stress, and anxiety
  disorders} {The neurocircuitry of fear, stress, and anxiety
  disorders}.{\BBCQ}
\newblock
\APACjournalVolNumPages{Neuropsychopharmacology}{35}{1}{169--191}.
\PrintBackRefs{\CurrentBib}

\bibitem [\protect \citeauthoryear {%
Siegrist%
\ \BBA {} Zingg%
}{%
Siegrist%
\ \BBA {} Zingg%
}{%
{\protect \APACyear {2014}}%
}]{%
siegrist2014role}
\APACinsertmetastar {%
siegrist2014role}%
\begin{APACrefauthors}%
Siegrist, M.%
\BCBT {}\ \BBA {} Zingg, A.%
\end{APACrefauthors}%
\unskip\
\newblock
\APACrefYearMonthDay{2014}{}{}.
\newblock
{\BBOQ}\APACrefatitle {The role of public trust during pandemics} {The role of
  public trust during pandemics}.{\BBCQ}
\newblock
\APACjournalVolNumPages{European psychologist}{}{}{}.
\PrintBackRefs{\CurrentBib}

\bibitem [\protect \citeauthoryear {%
Van~Bavel%
\ \protect \BOthers {.}}{%
Van~Bavel%
\ \protect \BOthers {.}}{%
{\protect \APACyear {2020}}%
}]{%
van2020using}
\APACinsertmetastar {%
van2020using}%
\begin{APACrefauthors}%
Van~Bavel, J\BPBI J.%
, Boggio, P.%
, Capraro, V.%
, Cichocka, A.%
, Cikara, M.%
, Crockett, M.%
\BDBL {}others%
\end{APACrefauthors}%
\unskip\
\newblock
\APACrefYearMonthDay{2020}{}{}.
\newblock
{\BBOQ}\APACrefatitle {Using social and behavioural science to support COVID-19
  pandemic response} {Using social and behavioural science to support covid-19
  pandemic response}.{\BBCQ}
\newblock

\PrintBackRefs{\CurrentBib}

\bibitem [\protect \citeauthoryear {%
Vaughan%
\ \BBA {} Tinker%
}{%
Vaughan%
\ \BBA {} Tinker%
}{%
{\protect \APACyear {2009}}%
}]{%
vaughan2009effective}
\APACinsertmetastar {%
vaughan2009effective}%
\begin{APACrefauthors}%
Vaughan, E.%
\BCBT {}\ \BBA {} Tinker, T.%
\end{APACrefauthors}%
\unskip\
\newblock
\APACrefYearMonthDay{2009}{}{}.
\newblock
{\BBOQ}\APACrefatitle {Effective health risk communication about pandemic
  influenza for vulnerable populations} {Effective health risk communication
  about pandemic influenza for vulnerable populations}.{\BBCQ}
\newblock
\APACjournalVolNumPages{American Journal of Public Health}{99}{S2}{S324--S332}.
\PrintBackRefs{\CurrentBib}

\bibitem [\protect \citeauthoryear {%
WHO%
}{%
WHO%
}{%
{\protect \APACyear {2018}}%
}]{%
Whoburden}
\APACinsertmetastar {%
Whoburden}%
\begin{APACrefauthors}%
WHO.%
\end{APACrefauthors}%
\unskip\
\newblock
\APACrefYearMonthDay{2018}{}{}.
\newblock
{\BBOQ}\APACrefatitle {Global Health Estimates 2016: Disease burden by Cause,
  Age, Sex, by Country and by Region, 2000-2016} {Global health estimates 2016:
  Disease burden by cause, age, sex, by country and by region,
  2000-2016}.{\BBCQ}
\newblock

\PrintBackRefs{\CurrentBib}

\bibitem [\protect \citeauthoryear {%
Zhao%
\ \BBA {} Xu%
}{%
Zhao%
\ \BBA {} Xu%
}{%
{\protect \APACyear {2020}}%
}]{%
zhao2020chinese}
\APACinsertmetastar {%
zhao2020chinese}%
\begin{APACrefauthors}%
Zhao, Y.%
\BCBT {}\ \BBA {} Xu, H.%
\end{APACrefauthors}%
\unskip\
\newblock
\APACrefYearMonthDay{2020}{}{}.
\newblock
{\BBOQ}\APACrefatitle {Chinese public attention to COVID-19 epidemic: Based on
  social media} {Chinese public attention to covid-19 epidemic: Based on social
  media}.{\BBCQ}
\newblock
\APACjournalVolNumPages{medRxiv}{}{}{}.
\PrintBackRefs{\CurrentBib}

\end{thebibliography}

\end{document}